\documentstyle[seceq,epsf]{ptptex}
\title{
Numerical Analysis of the Wave Function\\
 of the Multidimensional Universe
}

\author{
Hirotaka {\sc  Ochiai}\footnote{E-mail: ochiai@utap.phys.s.u-tokyo.ac.jp} 
and Katsuhiko {\sc Sato}\footnote{E-mail: sato@utap.phys.s.u-tokyo.ac.jp}
}

\inst{
$^{1)}$Department of Physics, School of Science, the University of Tokyo,\\
Tokyo 113-0033, Japan
\\
$^{2)}$Research Center for the Early Universe, School of Science,\\ 
the University of Tokyo, Tokyo 113-0033, Japan}
\recdate{
}
\abst{
In the framework of the Hartle-Hawking no-boundary proposal,
we investigate quantum creation of the multidimensional universe
with the cosmological constant $\Lambda$ but without matter fields.
In this paper we solved the Wheeler-de Witt equation numerically.
We find that the universe
in which both of the spaces expand exponentially
is the most probable in this model.
}
\begin{document}
\maketitle
\section{Introduction}
\label{sec:introduction}

In modern theories of unified physical interactions,
e.g. superstring theory and M-theory, 
spacetime has more than four dimensions. 
There are many works
on multidimensional cosmological models
(e.g. Refs. 1)--6) and references therein).
Some of them ~\cite{wu,huw,wu1,mel,is} are studied 
in the framework of the scenario of quantum creation of the universe.
~\cite{vilenkin,hh}
In the semi-classical approximation,
the effect of classical Euclidean solutions is essential,
and the wave function is estimated by
the Euclidean action.
However, a detailed analysis of the Euclidean Einstein equations
in the multidimensional cosmological model 
remained to be done.
In a previous paper,~\cite{os} we investigated quantum creation 
of the multidimensional universe
with the cosmological constant $\Lambda$ but without matter fields
by the method of the instanton.
In that paper, the metric was assumed 
to take the extended Friedmann-Robertson-Walker form.
We found that there are three instanton solutions
in the case that both the curvatures of the 
external and internal spaces are positive 
and $\Lambda>0$.
Furthermore, we found that 
the crossing point of three instanton solutions is 
a ``quasi-attractor'' point
from numerical analysis of the Euclidean solutions. 
Here, ``quasi-attractor'' point is a point to which 
most trajectories in the $a$ - $b$ plane circulate,
where $a$ and $b$ are the scale factors
of the external and internal spaces. 

We found that there are Lorentzian solutions which start near the
quasi-attractor and
three exact Lorentzian solutions
corresponding to three instanton solutions in Ref. 14).
Furthermore, from analyzing the classical solutions of the Einstein equations,
we found that it is probable that three universes are nucleated,
corresponding to these three solutions.
The first solution represents the case 
in which the external space expands and the internal space contracts.
The second solution represents the case in which
 both of the spaces expand exponentially.
The third solution represents the case
in which the internal space expands and the external space contracts.
In particular, the first solution describes the situation in which
the internal space remains microscopic and the external space evolves
to macroscopic our universe.
The goal of this paper is to clarify 
which solution is the most probable
by calculating the wave function of the universe.

The plan of the paper is as follows.
In $\S$2, we introduce our multidimensional cosmological model.
In $\S$3, we derive the Wheeler-de Witt equation
and solve it numerically by the difference method.
The wave function of the universe 
in Hartle-Hawking no-boundary proposal is then given.
Section 4 is devoted to a conclusion and remarks.
\section{The model}
We consider a $D=1+m+n$-dimensional vacuum universe.
Our assumed form for the metric is
\begin{equation}
d^2 s=g_{AB}dx^A dx^B=
-dt^2
+\exp{[2\beta^{0}(t)]}d\Omega_m^2
+\exp{[2\beta^{1}(t)]}d\Omega_n^2.
\end{equation}
The variables $\exp{[\beta^{0}(t)]}$ and $\exp{[\beta^{1}(t)]}$
are the scale factors of $m$-dimensional
and $n$-dimensional spaces, respectively.
The metrics $d\Omega_m^2$ and $d\Omega_n^2$
are these of an $m$-sphere and $n$-sphere with unit radius.
If the $m$-dimensional universe is our observed space,
then $m$=3.
The gravitational action is described as
\begin{equation}
S=\frac{1}{16\pi G}\int d^{D}x \sqrt{-g}
[R-2\Lambda]+S_{boundary},
\end{equation}
where $G$ is the $D$-dimensional gravitational constant, 
$\Lambda$ is the cosmological constant and
$g$ is the determinant of $g_{AB}$.
The last term $S_{boundary}$ 
is the York-Gibbons-Hawking boundary term.~\cite{york,gh}

Substituting the metric (2.1), the action is given by
\begin{equation}
S=\int dt \mu\exp(\gamma_0)\biggl[\frac{1}{2}G_{ij}\dot{\beta}^i\dot{\beta}^j
-V\biggr],
\end{equation}
where $i$, $j$$=$$0, 1$.
The minisupermetric $G_{ij}$ is defined as
\begin{equation}
G_{ij}=
\ \left(
\begin{array}{cc}
m-m^2&-mn\\
-mn&n-n^2\\
\end{array}
\right),\
\end{equation}
and the potential $V$ is defined as
\begin{equation}
V=-\frac{1}{2}m(m-1)k_m\exp(-2\beta^0)
-\frac{1}{2}n(n-1)k_n\exp(-2\beta^1)+\Lambda,
\end{equation}
where $k_m$ and $k_n$ denote the signs of the curvature of
the external and internal spaces, respectively.
As discussed in the Introduction
we consider the case of positive curvature ($k_m=k_n=1$).
Here the variable $\gamma_0$ is defined as
\begin{equation}
\gamma_0\equiv m\beta^{0}+n\beta^{1},
\end{equation}
and the constant $\mu$ is defined as
\begin{equation}
\mu=\frac{V_n}{16\pi G},
\end{equation}
where $V_n$ denotes the volume of the $n$-dimensional space.
The canonical conjugate momenta $\pi_i$ of the $\beta^{i}$ are given by
\begin{equation}
\pi_j=\mu\exp \gamma_0 G_{ij}\dot{\beta^i},
\end{equation}
where the inverse $G^{ij}$ of the minisupermetric is
\begin{equation}
G^{ij}=
\ \left(
\begin{array}{cc}
m-\frac{1}{D-2}&-\frac{1}{D-2}\\
-\frac{1}{D-2}&n-\frac{1}{D-2}\\
\end{array}
\right).\
\end{equation}
The Hamiltonian constraint is given by
\begin{equation}
H=\frac{1}{\mu}\exp (-\gamma_0)\biggl[\frac{1}{2}G^{ij}\pi_i\pi_j+
\mu^2\exp(2\gamma_0)V\biggr]
=0.
\end{equation}
\section{Numerical analysis of the Wheeler-De Witt equation}

In the canonical quantization procedure, the Hamiltonian constraint
leads to the Wheeler-De Witt equation.
In the Hawking-Page ansatz~\cite{hp}  
the Wheeler-De Witt equation is
\begin{equation}
\biggl(-\frac{1}{2}G^{ij}\partial_i\partial_j+U\biggr)\Psi=0,
\end{equation}
where $\partial_i=\frac{\partial}{\partial \beta^i}$.
Here, the new potential $U$ is defined as
\begin{equation}
U=\mu^2\exp(2\gamma_0)V.
\end{equation}
The inverse $G^{ij}$ of the minisupermetric is orthogonalized
under an appropriate linear transform of $\beta^{i}$.
For example, consider linear transform $\beta^{i}\rightarrow v^{i}$
defined by the following:
\begin{eqnarray}
v^0&=&\sqrt{\frac{m+n-1}{m+n}}(m\beta^{0}+n\beta^{1}),\\
v^1&=&\sqrt{\frac{mn}{(m+n)}}(-\beta^{0}+\beta^{1}).
\end{eqnarray}
Under this, the Wheeler-De Witt equation is transformed into
\begin{equation}
\biggl(\frac{1}{2}\frac{\partial}{\partial v^0}
\frac{\partial}{\partial v^0}-\frac{1}{2}\frac{\partial}{\partial v^1}
\frac{\partial}{\partial v^1}+U\biggr)\Psi=0.
\end{equation}
In order to obtain a convenient formulation for numerical calculations,
we further apply a rotation and use the following:
\begin{eqnarray}
z^0&=&\frac{1}{\sqrt{2}}(v^0-v^1),\\
z^1&=&\frac{1}{\sqrt{2}}(v^0+v^1).
\end{eqnarray}
Then the Wheeler-De Witt equation is given by
\begin{equation}
\biggl(\frac{\partial}{\partial z^0}
\frac{\partial}{\partial z^1}+U\biggr)\Psi=0.
\end{equation}

We solve the above Wheeler-De Witt equation using the difference method
following Hawking et al.~\cite{hms} and Hawking and Wu.\cite{hw}
The boundary condition is determined 
on the basis of the Hartle-Hawking proposal.
As the Euclidean action $S_E$ is vanishing for $z^0\rightarrow -\infty$
and $z^1\rightarrow -\infty$,
we use the approximation
$\Psi \sim e^{-S_E}\sim 1$
for $z^0\rightarrow -\infty$ and $z^1\rightarrow -\infty$.
We set the boundary condition as
$\Psi=1$, where $z^0$ and $z^1$ are small in relation to the curve $V=0$
for the numerical calculation.
Though we calculated the wave function with a different 
position of the boundary,
the behavior of the wave function was almost the same in the two cases.
Because the value of the constant $\mu$ is
independent of the dynamical properties,
we can set $\mu=1$ without loss of generality.

The wave function in our model
increases exponentially in the quantum region $U>0$
and is oscillatory in the classical region
$U<0$, as found in Ref. 18).
As a typical example,
we study the wave function in the case
$m=n=2$, $k_m=k_n=1$ and $\Lambda=1$.
The wave function increases exponentially 
near the boundaries $z^0=0$ and $z^1=0$.
It is shaped like a series of the mountains
along the curve where the potential $V$ is vanishing (see Fig. 1).
Furthermore, the wave function oscillates behind the mountains
and propagates along the line $z^0=z^1$ (see Fig. 2).
The wave function in the case 
$m=3$, $n=2$, $k_m=k_n=1$ and $\Lambda=1$ 
is displayed in Fig. 3.
From Figs. 2, 3, it is seen that
the wave propagates to 45 degrees in the $z^0-z^1$ plane,
independently of the number of spatial dimensions.
The path of the wave
corresponds to the solution 
that we found in Ref. 14):
\begin{eqnarray}
e^{\beta^0}&=&\sqrt{\frac{(m+n)(m-1)}{2\Lambda}}
\cosh\biggl(\sqrt{\frac{2\Lambda}{(m+n)(m+n-1)}}t\biggr),\\
e^{\beta^1}&=&\sqrt{\frac{(m+n)(n-1)}{2\Lambda}}
\cosh\biggl(\sqrt{\frac{2\Lambda}{(m+n)(m+n-1)}}t\biggr).
\end{eqnarray}

\begin{figure}
  \epsfxsize=76mm
  \centerline{\epsfbox{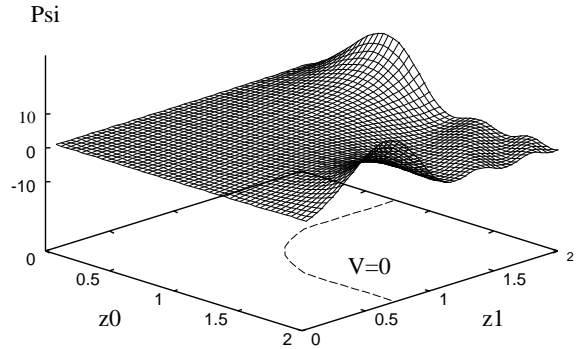}}
\caption{The wave function of the universe in the case
$m=n=2$, $k_m=k_n=1$ and $\Lambda=1$.
The boundary condition is $\Psi=1$ at $z^0=z^1=0$. 
}
\label{fig:1}
\end{figure}

\begin{figure}
  \epsfxsize=76mm
  \centerline{\epsfbox{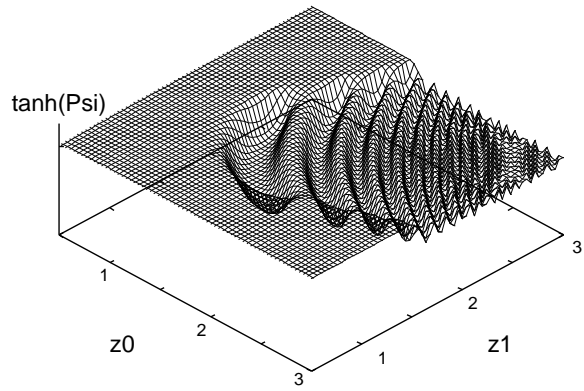}}
\caption{The wave function of the universe in the case
$m=n=2$, $k_m=k_n=1$ and $\Lambda=1$.
The boundary condition is $\Psi=1$ at $z^0=z^1=0$. 
Because the absolute value of the wave function grows exponentially,
$\tanh \Psi$ is shown.
}
\label{fig:2}
\end{figure}

\begin{figure}
  \epsfxsize=76mm
  \centerline{\epsfbox{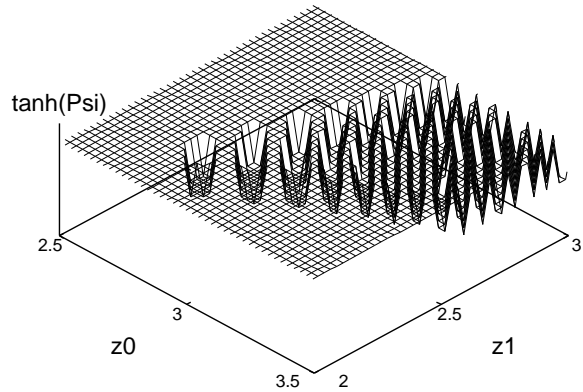}}
\caption{The wave function of the universe in the case
$m=3$, $n=2$, $k_m=k_n=1$ and $\Lambda=1$.
The boundary condition is that $\Psi=1$ at $z^0=z^1=0$. 
Because the absolute value of the wave function grows exponentially,
$\tanh \Psi$ is shown.
}
\label{fig:3}
\end{figure}

In Ref. 14),
we found three families of Lorentzian solutions
continued from the Euclidean solutions.
The first is the case 
that the external space expands and the internal space contracts.
The second is the case that both of the spaces expand exponentially.
The third is the case
that the internal space expands and the external space contracts.
As discussed in the Introduction,
it is of interest to clarify
which Lorentzian solution that we found 
in Ref. 14) is most probable.
We can determine this by interpreting the behavior of
the wave function.
To estimate Lorentzian solution,
we are interested in the classical region of the
wave function.
As in the case of Ref. 18),
the integral curves of $\nabla S$ correspond to
the trajectories of Lorentzian solutions in
the semi-classical approximation.
The classical trajectories are perpendicular to the wave crests
in the classical region.
Therefore,
the universe in which 
both external and internal spaces expand exponentially
[eqs. (3.10), (3.11)]
is most probable.
Unfortunately, this universe is not ours.
This most probable universe may be excluded
by the anthropic principle.
~\cite{ap}
If we do not depend on the anthropic principle, 
we must find a more realistic model which includes
a mechanism to exclude the expansion of the internal space.

\section{Summary and discussion}
\label{sec:summary and discussion}

We investigated the quantum creation of the multidimensional universe
with positive cosmological constant but without matter fields
in the framework of the Hartle-Hawking no-boundary proposal.
We calculated numerically the wave function of the universe in our model. 
The wave packet corresponds to the classical universe 
in which the
internal space and external space expand exponentially.
In conclusion, we found that the most probable universe is such
that both the external and internal spaces expand exponentially.
We can interpret that our universe corresponds to
the solution that the external space expands and the internal space
contracts by applying the anthropic principle.
However if we believe the most probable universe is our universe,
then a mechanism to exclude expansion of internal space is necessary.
For this purpose, it may be essential 
for the anisotropy of external space and internal space to be introduced.
A brane world scenario~\cite{hdd,rs}
studied recently is one approach that includes this.
\section*{Ackowledgements}

This work was supported in part 
by a Grant-in-Aid for Scientific Research (07CE2002) 
of the Ministry of Education, Science, Sports and Culture in Japan.


\end{document}